\documentclass[12pt, a4paper]{article}
\usepackage[a4paper, hmargin=2.5cm, vmargin=2cm]{geometry}
\usepackage{authblk}

\usepackage{amssymb}
\usepackage{amsmath}
\usepackage{stmaryrd}       
\usepackage{mathcomp}
\usepackage{color}
\usepackage{graphicx}
\usepackage{setspace}
\usepackage[colorinlistoftodos,prependcaption,textsize=footnotesize]{todonotes}
\usepackage{verbatim}		
\usepackage{courier}
\usepackage{mdframed}
\usepackage{longtable}
\usepackage{multirow}
\usepackage{caption}        
\usepackage{subcaption}
\usepackage{commath}        
\usepackage{epstopdf}
\usepackage{float}
\usepackage{bm}
\usepackage{textpos}
\usepackage{url}            
\usepackage[english]{babel}
\usepackage{dsfont}
\usepackage{array}
\usepackage{latexsym}
\usepackage{mathtools}
\usepackage{afterpage}
\usepackage{algorithm}
\usepackage{algorithmic}
\usepackage{soul}
\usepackage{xargs}          
\usepackage{keyval}         
\usepackage{pgfkeys}        
\usepackage{fix-cm}
\usepackage{blindtext}      
\usepackage{pifont}         

\usepackage{tikz}

\usepackage{tabularx}       
\usepackage{tcolorbox}      
\usepackage{colortbl}       
\usepackage{array}          

\usepackage{lipsum}
\usepackage{enumitem}       

\newcommand{\notshow}[1]{}
\newcommand{\red}[1]{{#1}}

\definecolor{AliceBlue}{rgb}{0.94,0.97,1.00}
\definecolor{AntiqueWhite1}{rgb}{1.00,0.94,0.86}
\definecolor{AntiqueWhite2}{rgb}{0.93,0.87,0.80}
\definecolor{AntiqueWhite3}{rgb}{0.80,0.75,0.69}
\definecolor{AntiqueWhite4}{rgb}{0.55,0.51,0.47}
\definecolor{AntiqueWhite}{rgb}{0.98,0.92,0.84}
\definecolor{BlanchedAlmond}{rgb}{1.00,0.92,0.80}
\definecolor{BlueViolet}{rgb}{0.54,0.17,0.89}
\definecolor{CadetBlue1}{rgb}{0.60,0.96,1.00}
\definecolor{CadetBlue2}{rgb}{0.56,0.90,0.93}
\definecolor{CadetBlue3}{rgb}{0.48,0.77,0.80}
\definecolor{CadetBlue4}{rgb}{0.33,0.53,0.55}
\definecolor{CadetBlue}{rgb}{0.37,0.62,0.63}
\definecolor{CornflowerBlue}{rgb}{0.39,0.58,0.93}
\definecolor{DarkBlue}{rgb}{0.00,0.00,0.55}
\definecolor{DarkCyan}{rgb}{0.00,0.55,0.55}
\definecolor{DarkGoldenrod1}{rgb}{1.00,0.73,0.06}
\definecolor{DarkGoldenrod2}{rgb}{0.93,0.68,0.05}
\definecolor{DarkGoldenrod3}{rgb}{0.80,0.58,0.05}
\definecolor{DarkGoldenrod4}{rgb}{0.55,0.40,0.03}
\definecolor{DarkGoldenrod}{rgb}{0.72,0.53,0.04}
\definecolor{DarkGray}{rgb}{0.66,0.66,0.66}
\definecolor{DarkGreen}{rgb}{0.00,0.39,0.00}
\definecolor{DarkGrey}{rgb}{0.66,0.66,0.66}
\definecolor{DarkKhaki}{rgb}{0.74,0.72,0.42}
\definecolor{DarkMagenta}{rgb}{0.55,0.00,0.55}
\definecolor{DarkOliveGreen1}{rgb}{0.79,1.00,0.44}
\definecolor{DarkOliveGreen2}{rgb}{0.74,0.93,0.41}
\definecolor{DarkOliveGreen3}{rgb}{0.64,0.80,0.35}
\definecolor{DarkOliveGreen4}{rgb}{0.43,0.55,0.24}
\definecolor{DarkOliveGreen}{rgb}{0.33,0.42,0.18}
\definecolor{DarkOrange1}{rgb}{1.00,0.50,0.00}
\definecolor{DarkOrange2}{rgb}{0.93,0.46,0.00}
\definecolor{DarkOrange3}{rgb}{0.80,0.40,0.00}
\definecolor{DarkOrange4}{rgb}{0.55,0.27,0.00}
\definecolor{DarkOrange}{rgb}{1.00,0.55,0.00}
\definecolor{DarkOrchid1}{rgb}{0.75,0.24,1.00}
\definecolor{DarkOrchid2}{rgb}{0.70,0.23,0.93}
\definecolor{DarkOrchid3}{rgb}{0.60,0.20,0.80}
\definecolor{DarkOrchid4}{rgb}{0.41,0.13,0.55}
\definecolor{DarkOrchid}{rgb}{0.60,0.20,0.80}
\definecolor{DarkRed}{rgb}{0.55,0.00,0.00}
\definecolor{DarkSalmon}{rgb}{0.91,0.59,0.48}
\definecolor{DarkSeaGreen1}{rgb}{0.76,1.00,0.76}
\definecolor{DarkSeaGreen2}{rgb}{0.71,0.93,0.71}
\definecolor{DarkSeaGreen3}{rgb}{0.61,0.80,0.61}
\definecolor{DarkSeaGreen4}{rgb}{0.41,0.55,0.41}
\definecolor{DarkSeaGreen}{rgb}{0.56,0.74,0.56}
\definecolor{DarkSlateBlue}{rgb}{0.28,0.24,0.55}
\definecolor{DarkSlateGray1}{rgb}{0.59,1.00,1.00}
\definecolor{DarkSlateGray2}{rgb}{0.55,0.93,0.93}
\definecolor{DarkSlateGray3}{rgb}{0.47,0.80,0.80}
\definecolor{DarkSlateGray4}{rgb}{0.32,0.55,0.55}
\definecolor{DarkSlateGray}{rgb}{0.18,0.31,0.31}
\definecolor{DarkSlateGrey}{rgb}{0.18,0.31,0.31}
\definecolor{DarkTurquoise}{rgb}{0.00,0.81,0.82}
\definecolor{DarkViolet}{rgb}{0.58,0.00,0.83}
\definecolor{DeepPink1}{rgb}{1.00,0.08,0.58}
\definecolor{DeepPink2}{rgb}{0.93,0.07,0.54}
\definecolor{DeepPink3}{rgb}{0.80,0.06,0.46}
\definecolor{DeepPink4}{rgb}{0.55,0.04,0.31}
\definecolor{DeepPink}{rgb}{1.00,0.08,0.58}
\definecolor{DeepSkyBlue1}{rgb}{0.00,0.75,1.00}
\definecolor{DeepSkyBlue2}{rgb}{0.00,0.70,0.93}
\definecolor{DeepSkyBlue3}{rgb}{0.00,0.60,0.80}
\definecolor{DeepSkyBlue4}{rgb}{0.00,0.41,0.55}
\definecolor{DeepSkyBlue}{rgb}{0.00,0.75,1.00}
\definecolor{DimGray}{rgb}{0.41,0.41,0.41}
\definecolor{DimGrey}{rgb}{0.41,0.41,0.41}
\definecolor{DodgerBlue1}{rgb}{0.12,0.56,1.00}
\definecolor{DodgerBlue2}{rgb}{0.11,0.53,0.93}
\definecolor{DodgerBlue3}{rgb}{0.09,0.45,0.80}
\definecolor{DodgerBlue4}{rgb}{0.06,0.31,0.55}
\definecolor{DodgerBlue}{rgb}{0.12,0.56,1.00}
\definecolor{FloralWhite}{rgb}{1.00,0.98,0.94}
\definecolor{ForestGreen}{rgb}{0.13,0.55,0.13}
\definecolor{GhostWhite}{rgb}{0.97,0.97,1.00}
\definecolor{GreenYellow}{rgb}{0.68,1.00,0.18}
\definecolor{HotPink1}{rgb}{1.00,0.43,0.71}
\definecolor{HotPink2}{rgb}{0.93,0.42,0.65}
\definecolor{HotPink3}{rgb}{0.80,0.38,0.56}
\definecolor{HotPink4}{rgb}{0.55,0.23,0.38}
\definecolor{HotPink}{rgb}{1.00,0.41,0.71}
\definecolor{IndianRed1}{rgb}{1.00,0.42,0.42}
\definecolor{IndianRed2}{rgb}{0.93,0.39,0.39}
\definecolor{IndianRed3}{rgb}{0.80,0.33,0.33}
\definecolor{IndianRed4}{rgb}{0.55,0.23,0.23}
\definecolor{IndianRed}{rgb}{0.80,0.36,0.36}
\definecolor{LavenderBlush1}{rgb}{1.00,0.94,0.96}
\definecolor{LavenderBlush2}{rgb}{0.93,0.88,0.90}
\definecolor{LavenderBlush3}{rgb}{0.80,0.76,0.77}
\definecolor{LavenderBlush4}{rgb}{0.55,0.51,0.53}
\definecolor{LavenderBlush}{rgb}{1.00,0.94,0.96}
\definecolor{LawnGreen}{rgb}{0.49,0.99,0.00}
\definecolor{LemonChiffon1}{rgb}{1.00,0.98,0.80}
\definecolor{LemonChiffon2}{rgb}{0.93,0.91,0.75}
\definecolor{LemonChiffon3}{rgb}{0.80,0.79,0.65}
\definecolor{LemonChiffon4}{rgb}{0.55,0.54,0.44}
\definecolor{LemonChiffon}{rgb}{1.00,0.98,0.80}
\definecolor{LightBlue1}{rgb}{0.75,0.94,1.00}
\definecolor{LightBlue2}{rgb}{0.70,0.87,0.93}
\definecolor{LightBlue3}{rgb}{0.60,0.75,0.80}
\definecolor{LightBlue4}{rgb}{0.41,0.51,0.55}
\definecolor{LightBlue}{rgb}{0.68,0.85,0.90}
\definecolor{LightCoral}{rgb}{0.94,0.50,0.50}
\definecolor{LightCyan1}{rgb}{0.88,1.00,1.00}
\definecolor{LightCyan2}{rgb}{0.82,0.93,0.93}
\definecolor{LightCyan3}{rgb}{0.71,0.80,0.80}
\definecolor{LightCyan4}{rgb}{0.48,0.55,0.55}
\definecolor{LightCyan}{rgb}{0.88,1.00,1.00}
\definecolor{LightGoldenrod1}{rgb}{1.00,0.93,0.55}
\definecolor{LightGoldenrod2}{rgb}{0.93,0.86,0.51}
\definecolor{LightGoldenrod3}{rgb}{0.80,0.75,0.44}
\definecolor{LightGoldenrod4}{rgb}{0.55,0.51,0.30}
\definecolor{LightGoldenrodYellow}{rgb}{0.98,0.98,0.82}
\definecolor{LightGoldenrod}{rgb}{0.93,0.87,0.51}
\definecolor{LightGray}{rgb}{0.83,0.83,0.83}
\definecolor{LightGreen}{rgb}{0.56,0.93,0.56}
\definecolor{LightGrey}{rgb}{0.83,0.83,0.83}
\definecolor{LightPink1}{rgb}{1.00,0.68,0.73}
\definecolor{LightPink2}{rgb}{0.93,0.64,0.68}
\definecolor{LightPink3}{rgb}{0.80,0.55,0.58}
\definecolor{LightPink4}{rgb}{0.55,0.37,0.40}
\definecolor{LightPink}{rgb}{1.00,0.71,0.76}
\definecolor{LightSalmon1}{rgb}{1.00,0.63,0.48}
\definecolor{LightSalmon2}{rgb}{0.93,0.58,0.45}
\definecolor{LightSalmon3}{rgb}{0.80,0.51,0.38}
\definecolor{LightSalmon4}{rgb}{0.55,0.34,0.26}
\definecolor{LightSalmon}{rgb}{1.00,0.63,0.48}
\definecolor{LightSeaGreen}{rgb}{0.13,0.70,0.67}
\definecolor{LightSkyBlue1}{rgb}{0.69,0.89,1.00}
\definecolor{LightSkyBlue2}{rgb}{0.64,0.83,0.93}
\definecolor{LightSkyBlue3}{rgb}{0.55,0.71,0.80}
\definecolor{LightSkyBlue4}{rgb}{0.38,0.48,0.55}
\definecolor{LightSkyBlue}{rgb}{0.53,0.81,0.98}
\definecolor{LightSlateBlue}{rgb}{0.52,0.44,1.00}
\definecolor{LightSlateGray}{rgb}{0.47,0.53,0.60}
\definecolor{LightSlateGrey}{rgb}{0.47,0.53,0.60}
\definecolor{LightSteelBlue1}{rgb}{0.79,0.88,1.00}
\definecolor{LightSteelBlue2}{rgb}{0.74,0.82,0.93}
\definecolor{LightSteelBlue3}{rgb}{0.64,0.71,0.80}
\definecolor{LightSteelBlue4}{rgb}{0.43,0.48,0.55}
\definecolor{LightSteelBlue}{rgb}{0.69,0.77,0.87}
\definecolor{LightYellow1}{rgb}{1.00,1.00,0.88}
\definecolor{LightYellow2}{rgb}{0.93,0.93,0.82}
\definecolor{LightYellow3}{rgb}{0.80,0.80,0.71}
\definecolor{LightYellow4}{rgb}{0.55,0.55,0.48}
\definecolor{LightYellow}{rgb}{1.00,1.00,0.88}
\definecolor{LimeGreen}{rgb}{0.20,0.80,0.20}
\definecolor{MediumAquamarine}{rgb}{0.40,0.80,0.67}
\definecolor{MediumBlue}{rgb}{0.00,0.00,0.80}
\definecolor{MediumOrchid1}{rgb}{0.88,0.40,1.00}
\definecolor{MediumOrchid2}{rgb}{0.82,0.37,0.93}
\definecolor{MediumOrchid3}{rgb}{0.71,0.32,0.80}
\definecolor{MediumOrchid4}{rgb}{0.48,0.22,0.55}
\definecolor{MediumOrchid}{rgb}{0.73,0.33,0.83}
\definecolor{MediumPurple1}{rgb}{0.67,0.51,1.00}
\definecolor{MediumPurple2}{rgb}{0.62,0.47,0.93}
\definecolor{MediumPurple3}{rgb}{0.54,0.41,0.80}
\definecolor{MediumPurple4}{rgb}{0.36,0.28,0.55}
\definecolor{MediumPurple}{rgb}{0.58,0.44,0.86}
\definecolor{MediumSeaGreen}{rgb}{0.24,0.70,0.44}
\definecolor{MediumSlateBlue}{rgb}{0.48,0.41,0.93}
\definecolor{MediumSpringGreen}{rgb}{0.00,0.98,0.60}
\definecolor{MediumTurquoise}{rgb}{0.28,0.82,0.80}
\definecolor{MediumVioletRed}{rgb}{0.78,0.08,0.52}
\definecolor{MidnightBlue}{rgb}{0.10,0.10,0.44}
\definecolor{MintCream}{rgb}{0.96,1.00,0.98}
\definecolor{MistyRose1}{rgb}{1.00,0.89,0.88}
\definecolor{MistyRose2}{rgb}{0.93,0.84,0.82}
\definecolor{MistyRose3}{rgb}{0.80,0.72,0.71}
\definecolor{MistyRose4}{rgb}{0.55,0.49,0.48}
\definecolor{MistyRose}{rgb}{1.00,0.89,0.88}
\definecolor{NavajoWhite1}{rgb}{1.00,0.87,0.68}
\definecolor{NavajoWhite2}{rgb}{0.93,0.81,0.63}
\definecolor{NavajoWhite3}{rgb}{0.80,0.70,0.55}
\definecolor{NavajoWhite4}{rgb}{0.55,0.47,0.37}
\definecolor{NavajoWhite}{rgb}{1.00,0.87,0.68}
\definecolor{NavyBlue}{rgb}{0.00,0.00,0.50}
\definecolor{OldLace}{rgb}{0.99,0.96,0.90}
\definecolor{OliveDrab1}{rgb}{0.75,1.00,0.24}
\definecolor{OliveDrab2}{rgb}{0.70,0.93,0.23}
\definecolor{OliveDrab3}{rgb}{0.60,0.80,0.20}
\definecolor{OliveDrab4}{rgb}{0.41,0.55,0.13}
\definecolor{OliveDrab}{rgb}{0.42,0.56,0.14}
\definecolor{OrangeRed1}{rgb}{1.00,0.27,0.00}
\definecolor{OrangeRed2}{rgb}{0.93,0.25,0.00}
\definecolor{OrangeRed3}{rgb}{0.80,0.22,0.00}
\definecolor{OrangeRed4}{rgb}{0.55,0.15,0.00}
\definecolor{OrangeRed}{rgb}{1.00,0.27,0.00}
\definecolor{PaleGoldenrod}{rgb}{0.93,0.91,0.67}
\definecolor{PaleGreen1}{rgb}{0.60,1.00,0.60}
\definecolor{PaleGreen2}{rgb}{0.56,0.93,0.56}
\definecolor{PaleGreen3}{rgb}{0.49,0.80,0.49}
\definecolor{PaleGreen4}{rgb}{0.33,0.55,0.33}
\definecolor{PaleGreen}{rgb}{0.60,0.98,0.60}
\definecolor{PaleTurquoise1}{rgb}{0.73,1.00,1.00}
\definecolor{PaleTurquoise2}{rgb}{0.68,0.93,0.93}
\definecolor{PaleTurquoise3}{rgb}{0.59,0.80,0.80}
\definecolor{PaleTurquoise4}{rgb}{0.40,0.55,0.55}
\definecolor{PaleTurquoise}{rgb}{0.69,0.93,0.93}
\definecolor{PaleVioletRed1}{rgb}{1.00,0.51,0.67}
\definecolor{PaleVioletRed2}{rgb}{0.93,0.47,0.62}
\definecolor{PaleVioletRed3}{rgb}{0.80,0.41,0.54}
\definecolor{PaleVioletRed4}{rgb}{0.55,0.28,0.36}
\definecolor{PaleVioletRed}{rgb}{0.86,0.44,0.58}
\definecolor{PapayaWhip}{rgb}{1.00,0.94,0.84}
\definecolor{PeachPuff1}{rgb}{1.00,0.85,0.73}
\definecolor{PeachPuff2}{rgb}{0.93,0.80,0.68}
\definecolor{PeachPuff3}{rgb}{0.80,0.69,0.58}
\definecolor{PeachPuff4}{rgb}{0.55,0.47,0.40}
\definecolor{PeachPuff}{rgb}{1.00,0.85,0.73}
\definecolor{PowderBlue}{rgb}{0.69,0.88,0.90}
\definecolor{RosyBrown1}{rgb}{1.00,0.76,0.76}
\definecolor{RosyBrown2}{rgb}{0.93,0.71,0.71}
\definecolor{RosyBrown3}{rgb}{0.80,0.61,0.61}
\definecolor{RosyBrown4}{rgb}{0.55,0.41,0.41}
\definecolor{RosyBrown}{rgb}{0.74,0.56,0.56}
\definecolor{RoyalBlue1}{rgb}{0.28,0.46,1.00}
\definecolor{RoyalBlue2}{rgb}{0.26,0.43,0.93}
\definecolor{RoyalBlue3}{rgb}{0.23,0.37,0.80}
\definecolor{RoyalBlue4}{rgb}{0.15,0.25,0.55}
\definecolor{RoyalBlue}{rgb}{0.25,0.41,0.88}
\definecolor{SaddleBrown}{rgb}{0.55,0.27,0.07}
\definecolor{SandyBrown}{rgb}{0.96,0.64,0.38}
\definecolor{SeaGreen1}{rgb}{0.33,1.00,0.62}
\definecolor{SeaGreen2}{rgb}{0.31,0.93,0.58}
\definecolor{SeaGreen3}{rgb}{0.26,0.80,0.50}
\definecolor{SeaGreen4}{rgb}{0.18,0.55,0.34}
\definecolor{SeaGreen}{rgb}{0.18,0.55,0.34}
\definecolor{SkyBlue1}{rgb}{0.53,0.81,1.00}
\definecolor{SkyBlue2}{rgb}{0.49,0.75,0.93}
\definecolor{SkyBlue3}{rgb}{0.42,0.65,0.80}
\definecolor{SkyBlue4}{rgb}{0.29,0.44,0.55}
\definecolor{SkyBlue}{rgb}{0.53,0.81,0.92}
\definecolor{SlateBlue1}{rgb}{0.51,0.44,1.00}
\definecolor{SlateBlue2}{rgb}{0.48,0.40,0.93}
\definecolor{SlateBlue3}{rgb}{0.41,0.35,0.80}
\definecolor{SlateBlue4}{rgb}{0.28,0.24,0.55}
\definecolor{SlateBlue}{rgb}{0.42,0.35,0.80}
\definecolor{SlateGray1}{rgb}{0.78,0.89,1.00}
\definecolor{SlateGray2}{rgb}{0.73,0.83,0.93}
\definecolor{SlateGray3}{rgb}{0.62,0.71,0.80}
\definecolor{SlateGray4}{rgb}{0.42,0.48,0.55}
\definecolor{SlateGray}{rgb}{0.44,0.50,0.56}
\definecolor{SlateGrey}{rgb}{0.44,0.50,0.56}
\definecolor{SpringGreen1}{rgb}{0.00,1.00,0.50}
\definecolor{SpringGreen2}{rgb}{0.00,0.93,0.46}
\definecolor{SpringGreen3}{rgb}{0.00,0.80,0.40}
\definecolor{SpringGreen4}{rgb}{0.00,0.55,0.27}
\definecolor{SpringGreen}{rgb}{0.00,1.00,0.50}
\definecolor{SteelBlue1}{rgb}{0.39,0.72,1.00}
\definecolor{SteelBlue2}{rgb}{0.36,0.67,0.93}
\definecolor{SteelBlue3}{rgb}{0.31,0.58,0.80}
\definecolor{SteelBlue4}{rgb}{0.21,0.39,0.55}
\definecolor{SteelBlue}{rgb}{0.27,0.51,0.71}
\definecolor{VioletRed1}{rgb}{1.00,0.24,0.59}
\definecolor{VioletRed2}{rgb}{0.93,0.23,0.55}
\definecolor{VioletRed3}{rgb}{0.80,0.20,0.47}
\definecolor{VioletRed4}{rgb}{0.55,0.13,0.32}
\definecolor{VioletRed}{rgb}{0.82,0.13,0.56}
\definecolor{WhiteSmoke}{rgb}{0.96,0.96,0.96}
\definecolor{YellowGreen}{rgb}{0.60,0.80,0.20}
\definecolor{aliceblue}{rgb}{0.94,0.97,1.00}
\definecolor{antiquewhite}{rgb}{0.98,0.92,0.84}
\definecolor{aquamarine1}{rgb}{0.50,1.00,0.83}
\definecolor{aquamarine2}{rgb}{0.46,0.93,0.78}
\definecolor{aquamarine3}{rgb}{0.40,0.80,0.67}
\definecolor{aquamarine4}{rgb}{0.27,0.55,0.45}
\definecolor{aquamarine}{rgb}{0.50,1.00,0.83}
\definecolor{azure1}{rgb}{0.94,1.00,1.00}
\definecolor{azure2}{rgb}{0.88,0.93,0.93}
\definecolor{azure3}{rgb}{0.76,0.80,0.80}
\definecolor{azure4}{rgb}{0.51,0.55,0.55}
\definecolor{azure}{rgb}{0.94,1.00,1.00}
\definecolor{beige}{rgb}{0.96,0.96,0.86}
\definecolor{bisque1}{rgb}{1.00,0.89,0.77}
\definecolor{bisque2}{rgb}{0.93,0.84,0.72}
\definecolor{bisque3}{rgb}{0.80,0.72,0.62}
\definecolor{bisque4}{rgb}{0.55,0.49,0.42}
\definecolor{bisque}{rgb}{1.00,0.89,0.77}
\definecolor{black}{rgb}{0.00,0.00,0.00}
\definecolor{blanchedalmond}{rgb}{1.00,0.92,0.80}
\definecolor{blue1}{rgb}{0.00,0.00,1.00}
\definecolor{blue2}{rgb}{0.00,0.00,0.93}
\definecolor{blue3}{rgb}{0.00,0.00,0.80}
\definecolor{blue4}{rgb}{0.00,0.00,0.55}
\definecolor{blueviolet}{rgb}{0.54,0.17,0.89}
\definecolor{blue}{rgb}{0.00,0.00,1.00}
\definecolor{brown1}{rgb}{1.00,0.25,0.25}
\definecolor{brown2}{rgb}{0.93,0.23,0.23}
\definecolor{brown3}{rgb}{0.80,0.20,0.20}
\definecolor{brown4}{rgb}{0.55,0.14,0.14}
\definecolor{brown}{rgb}{0.65,0.16,0.16}
\definecolor{burlywood1}{rgb}{1.00,0.83,0.61}
\definecolor{burlywood2}{rgb}{0.93,0.77,0.57}
\definecolor{burlywood3}{rgb}{0.80,0.67,0.49}
\definecolor{burlywood4}{rgb}{0.55,0.45,0.33}
\definecolor{burlywood}{rgb}{0.87,0.72,0.53}
\definecolor{cadetblue}{rgb}{0.37,0.62,0.63}
\definecolor{chartreuse1}{rgb}{0.50,1.00,0.00}
\definecolor{chartreuse2}{rgb}{0.46,0.93,0.00}
\definecolor{chartreuse3}{rgb}{0.40,0.80,0.00}
\definecolor{chartreuse4}{rgb}{0.27,0.55,0.00}
\definecolor{chartreuse}{rgb}{0.50,1.00,0.00}
\definecolor{chocolate1}{rgb}{1.00,0.50,0.14}
\definecolor{chocolate2}{rgb}{0.93,0.46,0.13}
\definecolor{chocolate3}{rgb}{0.80,0.40,0.11}
\definecolor{chocolate4}{rgb}{0.55,0.27,0.07}
\definecolor{chocolate}{rgb}{0.82,0.41,0.12}
\definecolor{coral1}{rgb}{1.00,0.45,0.34}
\definecolor{coral2}{rgb}{0.93,0.42,0.31}
\definecolor{coral3}{rgb}{0.80,0.36,0.27}
\definecolor{coral4}{rgb}{0.55,0.24,0.18}
\definecolor{coral}{rgb}{1.00,0.50,0.31}
\definecolor{cornflowerblue}{rgb}{0.39,0.58,0.93}
\definecolor{cornsilk1}{rgb}{1.00,0.97,0.86}
\definecolor{cornsilk2}{rgb}{0.93,0.91,0.80}
\definecolor{cornsilk3}{rgb}{0.80,0.78,0.69}
\definecolor{cornsilk4}{rgb}{0.55,0.53,0.47}
\definecolor{cornsilk}{rgb}{1.00,0.97,0.86}
\definecolor{cyan1}{rgb}{0.00,1.00,1.00}
\definecolor{cyan2}{rgb}{0.00,0.93,0.93}
\definecolor{cyan3}{rgb}{0.00,0.80,0.80}
\definecolor{cyan4}{rgb}{0.00,0.55,0.55}
\definecolor{cyan}{rgb}{0.00,1.00,1.00}
\definecolor{darkblue}{rgb}{0.00,0.00,0.55}
\definecolor{darkcyan}{rgb}{0.00,0.55,0.55}
\definecolor{darkgoldenrod}{rgb}{0.72,0.53,0.04}
\definecolor{darkgray}{rgb}{0.66,0.66,0.66}
\definecolor{darkgreen}{rgb}{0.00,0.39,0.00}
\definecolor{darkgrey}{rgb}{0.66,0.66,0.66}
\definecolor{darkkhaki}{rgb}{0.74,0.72,0.42}
\definecolor{darkmagenta}{rgb}{0.55,0.00,0.55}
\definecolor{darkolive}{rgb}{0.33,0.42,0.18}
\definecolor{darkorange}{rgb}{1.00,0.55,0.00}
\definecolor{darkorchid}{rgb}{0.60,0.20,0.80}
\definecolor{darkred}{rgb}{0.55,0.00,0.00}
\definecolor{darksalmon}{rgb}{0.91,0.59,0.48}
\definecolor{darksea}{rgb}{0.56,0.74,0.56}
\definecolor{darkslate}{rgb}{0.18,0.31,0.31}
\definecolor{darkslate}{rgb}{0.18,0.31,0.31}
\definecolor{darkslate}{rgb}{0.28,0.24,0.55}
\definecolor{darkturquoise}{rgb}{0.00,0.81,0.82}
\definecolor{darkviolet}{rgb}{0.58,0.00,0.83}
\definecolor{deeppink}{rgb}{1.00,0.08,0.58}
\definecolor{deepsky}{rgb}{0.00,0.75,1.00}
\definecolor{dimgray}{rgb}{0.41,0.41,0.41}
\definecolor{dimgrey}{rgb}{0.41,0.41,0.41}
\definecolor{dodgerblue}{rgb}{0.12,0.56,1.00}
\definecolor{firebrick1}{rgb}{1.00,0.19,0.19}
\definecolor{firebrick2}{rgb}{0.93,0.17,0.17}
\definecolor{firebrick3}{rgb}{0.80,0.15,0.15}
\definecolor{firebrick4}{rgb}{0.55,0.10,0.10}
\definecolor{firebrick}{rgb}{0.70,0.13,0.13}
\definecolor{floralwhite}{rgb}{1.00,0.98,0.94}
\definecolor{forestgreen}{rgb}{0.13,0.55,0.13}
\definecolor{gainsboro}{rgb}{0.86,0.86,0.86}
\definecolor{ghostwhite}{rgb}{0.97,0.97,1.00}
\definecolor{gold1}{rgb}{1.00,0.84,0.00}
\definecolor{gold2}{rgb}{0.93,0.79,0.00}
\definecolor{gold3}{rgb}{0.80,0.68,0.00}
\definecolor{gold4}{rgb}{0.55,0.46,0.00}
\definecolor{goldenrod1}{rgb}{1.00,0.76,0.15}
\definecolor{goldenrod2}{rgb}{0.93,0.71,0.13}
\definecolor{goldenrod3}{rgb}{0.80,0.61,0.11}
\definecolor{goldenrod4}{rgb}{0.55,0.41,0.08}
\definecolor{goldenrod}{rgb}{0.85,0.65,0.13}
\definecolor{gold}{rgb}{1.00,0.84,0.00}
\definecolor{gray0}{rgb}{0.00,0.00,0.00}
\definecolor{gray100}{rgb}{1.00,1.00,1.00}
\definecolor{gray10}{rgb}{0.10,0.10,0.10}
\definecolor{gray11}{rgb}{0.11,0.11,0.11}
\definecolor{gray12}{rgb}{0.12,0.12,0.12}
\definecolor{gray13}{rgb}{0.13,0.13,0.13}
\definecolor{gray14}{rgb}{0.14,0.14,0.14}
\definecolor{gray15}{rgb}{0.15,0.15,0.15}
\definecolor{gray16}{rgb}{0.16,0.16,0.16}
\definecolor{gray17}{rgb}{0.17,0.17,0.17}
\definecolor{gray18}{rgb}{0.18,0.18,0.18}
\definecolor{gray19}{rgb}{0.19,0.19,0.19}
\definecolor{gray1}{rgb}{0.01,0.01,0.01}
\definecolor{gray20}{rgb}{0.20,0.20,0.20}
\definecolor{gray21}{rgb}{0.21,0.21,0.21}
\definecolor{gray22}{rgb}{0.22,0.22,0.22}
\definecolor{gray23}{rgb}{0.23,0.23,0.23}
\definecolor{gray24}{rgb}{0.24,0.24,0.24}
\definecolor{gray25}{rgb}{0.25,0.25,0.25}
\definecolor{gray26}{rgb}{0.26,0.26,0.26}
\definecolor{gray27}{rgb}{0.27,0.27,0.27}
\definecolor{gray28}{rgb}{0.28,0.28,0.28}
\definecolor{gray29}{rgb}{0.29,0.29,0.29}
\definecolor{gray2}{rgb}{0.02,0.02,0.02}
\definecolor{gray30}{rgb}{0.30,0.30,0.30}
\definecolor{gray31}{rgb}{0.31,0.31,0.31}
\definecolor{gray32}{rgb}{0.32,0.32,0.32}
\definecolor{gray33}{rgb}{0.33,0.33,0.33}
\definecolor{gray34}{rgb}{0.34,0.34,0.34}
\definecolor{gray35}{rgb}{0.35,0.35,0.35}
\definecolor{gray36}{rgb}{0.36,0.36,0.36}
\definecolor{gray37}{rgb}{0.37,0.37,0.37}
\definecolor{gray38}{rgb}{0.38,0.38,0.38}
\definecolor{gray39}{rgb}{0.39,0.39,0.39}
\definecolor{gray3}{rgb}{0.03,0.03,0.03}
\definecolor{gray40}{rgb}{0.40,0.40,0.40}
\definecolor{gray41}{rgb}{0.41,0.41,0.41}
\definecolor{gray42}{rgb}{0.42,0.42,0.42}
\definecolor{gray43}{rgb}{0.43,0.43,0.43}
\definecolor{gray44}{rgb}{0.44,0.44,0.44}
\definecolor{gray45}{rgb}{0.45,0.45,0.45}
\definecolor{gray46}{rgb}{0.46,0.46,0.46}
\definecolor{gray47}{rgb}{0.47,0.47,0.47}
\definecolor{gray48}{rgb}{0.48,0.48,0.48}
\definecolor{gray49}{rgb}{0.49,0.49,0.49}
\definecolor{gray4}{rgb}{0.04,0.04,0.04}
\definecolor{gray50}{rgb}{0.50,0.50,0.50}
\definecolor{gray51}{rgb}{0.51,0.51,0.51}
\definecolor{gray52}{rgb}{0.52,0.52,0.52}
\definecolor{gray53}{rgb}{0.53,0.53,0.53}
\definecolor{gray54}{rgb}{0.54,0.54,0.54}
\definecolor{gray55}{rgb}{0.55,0.55,0.55}
\definecolor{gray56}{rgb}{0.56,0.56,0.56}
\definecolor{gray57}{rgb}{0.57,0.57,0.57}
\definecolor{gray58}{rgb}{0.58,0.58,0.58}
\definecolor{gray59}{rgb}{0.59,0.59,0.59}
\definecolor{gray5}{rgb}{0.05,0.05,0.05}
\definecolor{gray60}{rgb}{0.60,0.60,0.60}
\definecolor{gray61}{rgb}{0.61,0.61,0.61}
\definecolor{gray62}{rgb}{0.62,0.62,0.62}
\definecolor{gray63}{rgb}{0.63,0.63,0.63}
\definecolor{gray64}{rgb}{0.64,0.64,0.64}
\definecolor{gray65}{rgb}{0.65,0.65,0.65}
\definecolor{gray66}{rgb}{0.66,0.66,0.66}
\definecolor{gray67}{rgb}{0.67,0.67,0.67}
\definecolor{gray68}{rgb}{0.68,0.68,0.68}
\definecolor{gray69}{rgb}{0.69,0.69,0.69}
\definecolor{gray6}{rgb}{0.06,0.06,0.06}
\definecolor{gray70}{rgb}{0.70,0.70,0.70}
\definecolor{gray71}{rgb}{0.71,0.71,0.71}
\definecolor{gray72}{rgb}{0.72,0.72,0.72}
\definecolor{gray73}{rgb}{0.73,0.73,0.73}
\definecolor{gray74}{rgb}{0.74,0.74,0.74}
\definecolor{gray75}{rgb}{0.75,0.75,0.75}
\definecolor{gray76}{rgb}{0.76,0.76,0.76}
\definecolor{gray77}{rgb}{0.77,0.77,0.77}
\definecolor{gray78}{rgb}{0.78,0.78,0.78}
\definecolor{gray79}{rgb}{0.79,0.79,0.79}
\definecolor{gray7}{rgb}{0.07,0.07,0.07}
\definecolor{gray80}{rgb}{0.80,0.80,0.80}
\definecolor{gray81}{rgb}{0.81,0.81,0.81}
\definecolor{gray82}{rgb}{0.82,0.82,0.82}
\definecolor{gray83}{rgb}{0.83,0.83,0.83}
\definecolor{gray84}{rgb}{0.84,0.84,0.84}
\definecolor{gray85}{rgb}{0.85,0.85,0.85}
\definecolor{gray86}{rgb}{0.86,0.86,0.86}
\definecolor{gray87}{rgb}{0.87,0.87,0.87}
\definecolor{gray88}{rgb}{0.88,0.88,0.88}
\definecolor{gray89}{rgb}{0.89,0.89,0.89}
\definecolor{gray8}{rgb}{0.08,0.08,0.08}
\definecolor{gray90}{rgb}{0.90,0.90,0.90}
\definecolor{gray91}{rgb}{0.91,0.91,0.91}
\definecolor{gray92}{rgb}{0.92,0.92,0.92}
\definecolor{gray93}{rgb}{0.93,0.93,0.93}
\definecolor{gray94}{rgb}{0.94,0.94,0.94}
\definecolor{gray95}{rgb}{0.95,0.95,0.95}
\definecolor{gray96}{rgb}{0.96,0.96,0.96}
\definecolor{gray97}{rgb}{0.97,0.97,0.97}
\definecolor{gray98}{rgb}{0.98,0.98,0.98}
\definecolor{gray99}{rgb}{0.99,0.99,0.99}
\definecolor{gray9}{rgb}{0.09,0.09,0.09}
\definecolor{gray}{rgb}{0.75,0.75,0.75}
\definecolor{green1}{rgb}{0.00,1.00,0.00}
\definecolor{green2}{rgb}{0.00,0.93,0.00}
\definecolor{green3}{rgb}{0.00,0.80,0.00}
\definecolor{green4}{rgb}{0.00,0.55,0.00}
\definecolor{greenyellow}{rgb}{0.68,1.00,0.18}
\definecolor{green}{rgb}{0.00,1.00,0.00}
\definecolor{grey0}{rgb}{0.00,0.00,0.00}
\definecolor{grey100}{rgb}{1.00,1.00,1.00}
\definecolor{grey10}{rgb}{0.10,0.10,0.10}
\definecolor{grey11}{rgb}{0.11,0.11,0.11}
\definecolor{grey12}{rgb}{0.12,0.12,0.12}
\definecolor{grey13}{rgb}{0.13,0.13,0.13}
\definecolor{grey14}{rgb}{0.14,0.14,0.14}
\definecolor{grey15}{rgb}{0.15,0.15,0.15}
\definecolor{grey16}{rgb}{0.16,0.16,0.16}
\definecolor{grey17}{rgb}{0.17,0.17,0.17}
\definecolor{grey18}{rgb}{0.18,0.18,0.18}
\definecolor{grey19}{rgb}{0.19,0.19,0.19}
\definecolor{grey1}{rgb}{0.01,0.01,0.01}
\definecolor{grey20}{rgb}{0.20,0.20,0.20}
\definecolor{grey21}{rgb}{0.21,0.21,0.21}
\definecolor{grey22}{rgb}{0.22,0.22,0.22}
\definecolor{grey23}{rgb}{0.23,0.23,0.23}
\definecolor{grey24}{rgb}{0.24,0.24,0.24}
\definecolor{grey25}{rgb}{0.25,0.25,0.25}
\definecolor{grey26}{rgb}{0.26,0.26,0.26}
\definecolor{grey27}{rgb}{0.27,0.27,0.27}
\definecolor{grey28}{rgb}{0.28,0.28,0.28}
\definecolor{grey29}{rgb}{0.29,0.29,0.29}
\definecolor{grey2}{rgb}{0.02,0.02,0.02}
\definecolor{grey30}{rgb}{0.30,0.30,0.30}
\definecolor{grey31}{rgb}{0.31,0.31,0.31}
\definecolor{grey32}{rgb}{0.32,0.32,0.32}
\definecolor{grey33}{rgb}{0.33,0.33,0.33}
\definecolor{grey34}{rgb}{0.34,0.34,0.34}
\definecolor{grey35}{rgb}{0.35,0.35,0.35}
\definecolor{grey36}{rgb}{0.36,0.36,0.36}
\definecolor{grey37}{rgb}{0.37,0.37,0.37}
\definecolor{grey38}{rgb}{0.38,0.38,0.38}
\definecolor{grey39}{rgb}{0.39,0.39,0.39}
\definecolor{grey3}{rgb}{0.03,0.03,0.03}
\definecolor{grey40}{rgb}{0.40,0.40,0.40}
\definecolor{grey41}{rgb}{0.41,0.41,0.41}
\definecolor{grey42}{rgb}{0.42,0.42,0.42}
\definecolor{grey43}{rgb}{0.43,0.43,0.43}
\definecolor{grey44}{rgb}{0.44,0.44,0.44}
\definecolor{grey45}{rgb}{0.45,0.45,0.45}
\definecolor{grey46}{rgb}{0.46,0.46,0.46}
\definecolor{grey47}{rgb}{0.47,0.47,0.47}
\definecolor{grey48}{rgb}{0.48,0.48,0.48}
\definecolor{grey49}{rgb}{0.49,0.49,0.49}
\definecolor{grey4}{rgb}{0.04,0.04,0.04}
\definecolor{grey50}{rgb}{0.50,0.50,0.50}
\definecolor{grey51}{rgb}{0.51,0.51,0.51}
\definecolor{grey52}{rgb}{0.52,0.52,0.52}
\definecolor{grey53}{rgb}{0.53,0.53,0.53}
\definecolor{grey54}{rgb}{0.54,0.54,0.54}
\definecolor{grey55}{rgb}{0.55,0.55,0.55}
\definecolor{grey56}{rgb}{0.56,0.56,0.56}
\definecolor{grey57}{rgb}{0.57,0.57,0.57}
\definecolor{grey58}{rgb}{0.58,0.58,0.58}
\definecolor{grey59}{rgb}{0.59,0.59,0.59}
\definecolor{grey5}{rgb}{0.05,0.05,0.05}
\definecolor{grey60}{rgb}{0.60,0.60,0.60}
\definecolor{grey61}{rgb}{0.61,0.61,0.61}
\definecolor{grey62}{rgb}{0.62,0.62,0.62}
\definecolor{grey63}{rgb}{0.63,0.63,0.63}
\definecolor{grey64}{rgb}{0.64,0.64,0.64}
\definecolor{grey65}{rgb}{0.65,0.65,0.65}
\definecolor{grey66}{rgb}{0.66,0.66,0.66}
\definecolor{grey67}{rgb}{0.67,0.67,0.67}
\definecolor{grey68}{rgb}{0.68,0.68,0.68}
\definecolor{grey69}{rgb}{0.69,0.69,0.69}
\definecolor{grey6}{rgb}{0.06,0.06,0.06}
\definecolor{grey70}{rgb}{0.70,0.70,0.70}
\definecolor{grey71}{rgb}{0.71,0.71,0.71}
\definecolor{grey72}{rgb}{0.72,0.72,0.72}
\definecolor{grey73}{rgb}{0.73,0.73,0.73}
\definecolor{grey74}{rgb}{0.74,0.74,0.74}
\definecolor{grey75}{rgb}{0.75,0.75,0.75}
\definecolor{grey76}{rgb}{0.76,0.76,0.76}
\definecolor{grey77}{rgb}{0.77,0.77,0.77}
\definecolor{grey78}{rgb}{0.78,0.78,0.78}
\definecolor{grey79}{rgb}{0.79,0.79,0.79}
\definecolor{grey7}{rgb}{0.07,0.07,0.07}
\definecolor{grey80}{rgb}{0.80,0.80,0.80}
\definecolor{grey81}{rgb}{0.81,0.81,0.81}
\definecolor{grey82}{rgb}{0.82,0.82,0.82}
\definecolor{grey83}{rgb}{0.83,0.83,0.83}
\definecolor{grey84}{rgb}{0.84,0.84,0.84}
\definecolor{grey85}{rgb}{0.85,0.85,0.85}
\definecolor{grey86}{rgb}{0.86,0.86,0.86}
\definecolor{grey87}{rgb}{0.87,0.87,0.87}
\definecolor{grey88}{rgb}{0.88,0.88,0.88}
\definecolor{grey89}{rgb}{0.89,0.89,0.89}
\definecolor{grey8}{rgb}{0.08,0.08,0.08}
\definecolor{grey90}{rgb}{0.90,0.90,0.90}
\definecolor{grey91}{rgb}{0.91,0.91,0.91}
\definecolor{grey92}{rgb}{0.92,0.92,0.92}
\definecolor{grey93}{rgb}{0.93,0.93,0.93}
\definecolor{grey94}{rgb}{0.94,0.94,0.94}
\definecolor{grey95}{rgb}{0.95,0.95,0.95}
\definecolor{grey96}{rgb}{0.96,0.96,0.96}
\definecolor{grey97}{rgb}{0.97,0.97,0.97}
\definecolor{grey98}{rgb}{0.98,0.98,0.98}
\definecolor{grey99}{rgb}{0.99,0.99,0.99}
\definecolor{grey9}{rgb}{0.09,0.09,0.09}
\definecolor{grey}{rgb}{0.75,0.75,0.75}
\definecolor{honeydew1}{rgb}{0.94,1.00,0.94}
\definecolor{honeydew2}{rgb}{0.88,0.93,0.88}
\definecolor{honeydew3}{rgb}{0.76,0.80,0.76}
\definecolor{honeydew4}{rgb}{0.51,0.55,0.51}
\definecolor{honeydew}{rgb}{0.94,1.00,0.94}
\definecolor{hotpink}{rgb}{1.00,0.41,0.71}
\definecolor{indianred}{rgb}{0.80,0.36,0.36}
\definecolor{ivory1}{rgb}{1.00,1.00,0.94}
\definecolor{ivory2}{rgb}{0.93,0.93,0.88}
\definecolor{ivory3}{rgb}{0.80,0.80,0.76}
\definecolor{ivory4}{rgb}{0.55,0.55,0.51}
\definecolor{ivory}{rgb}{1.00,1.00,0.94}
\definecolor{khaki1}{rgb}{1.00,0.96,0.56}
\definecolor{khaki2}{rgb}{0.93,0.90,0.52}
\definecolor{khaki3}{rgb}{0.80,0.78,0.45}
\definecolor{khaki4}{rgb}{0.55,0.53,0.31}
\definecolor{khaki}{rgb}{0.94,0.90,0.55}
\definecolor{lavenderblush}{rgb}{1.00,0.94,0.96}
\definecolor{lavender}{rgb}{0.90,0.90,0.98}
\definecolor{lawngreen}{rgb}{0.49,0.99,0.00}
\definecolor{lemonchiffon}{rgb}{1.00,0.98,0.80}
\definecolor{lightblue}{rgb}{0.68,0.85,0.90}
\definecolor{lightcoral}{rgb}{0.94,0.50,0.50}
\definecolor{lightcyan}{rgb}{0.88,1.00,1.00}
\definecolor{lightgoldenrod}{rgb}{0.93,0.87,0.51}
\definecolor{lightgoldenrod}{rgb}{0.98,0.98,0.82}
\definecolor{lightgray}{rgb}{0.83,0.83,0.83}
\definecolor{lightgreen}{rgb}{0.56,0.93,0.56}
\definecolor{lightgrey}{rgb}{0.83,0.83,0.83}
\definecolor{lightpink}{rgb}{1.00,0.71,0.76}
\definecolor{lightsalmon}{rgb}{1.00,0.63,0.48}
\definecolor{lightsea}{rgb}{0.13,0.70,0.67}
\definecolor{lightsky}{rgb}{0.53,0.81,0.98}
\definecolor{lightslate}{rgb}{0.47,0.53,0.60}
\definecolor{lightslate}{rgb}{0.47,0.53,0.60}
\definecolor{lightslate}{rgb}{0.52,0.44,1.00}
\definecolor{lightsteel}{rgb}{0.69,0.77,0.87}
\definecolor{lightyellow}{rgb}{1.00,1.00,0.88}
\definecolor{limegreen}{rgb}{0.20,0.80,0.20}
\definecolor{linen}{rgb}{0.98,0.94,0.90}
\definecolor{magenta1}{rgb}{1.00,0.00,1.00}
\definecolor{magenta2}{rgb}{0.93,0.00,0.93}
\definecolor{magenta3}{rgb}{0.80,0.00,0.80}
\definecolor{magenta4}{rgb}{0.55,0.00,0.55}
\definecolor{magenta}{rgb}{1.00,0.00,1.00}
\definecolor{maroon1}{rgb}{1.00,0.20,0.70}
\definecolor{maroon2}{rgb}{0.93,0.19,0.65}
\definecolor{maroon3}{rgb}{0.80,0.16,0.56}
\definecolor{maroon4}{rgb}{0.55,0.11,0.38}
\definecolor{maroon}{rgb}{0.69,0.19,0.38}
\definecolor{mediumaquamarine}{rgb}{0.40,0.80,0.67}
\definecolor{mediumblue}{rgb}{0.00,0.00,0.80}
\definecolor{mediumorchid}{rgb}{0.73,0.33,0.83}
\definecolor{mediumpurple}{rgb}{0.58,0.44,0.86}
\definecolor{mediumsea}{rgb}{0.24,0.70,0.44}
\definecolor{mediumslate}{rgb}{0.48,0.41,0.93}
\definecolor{mediumspring}{rgb}{0.00,0.98,0.60}
\definecolor{mediumturquoise}{rgb}{0.28,0.82,0.80}
\definecolor{mediumviolet}{rgb}{0.78,0.08,0.52}
\definecolor{midnightblue}{rgb}{0.10,0.10,0.44}
\definecolor{mintcream}{rgb}{0.96,1.00,0.98}
\definecolor{mistyrose}{rgb}{1.00,0.89,0.88}
\definecolor{moccasin}{rgb}{1.00,0.89,0.71}
\definecolor{navajowhite}{rgb}{1.00,0.87,0.68}
\definecolor{navyblue}{rgb}{0.00,0.00,0.50}
\definecolor{navy}{rgb}{0.00,0.00,0.50}
\definecolor{oldlace}{rgb}{0.99,0.96,0.90}
\definecolor{olivedrab}{rgb}{0.42,0.56,0.14}
\definecolor{orange1}{rgb}{1.00,0.65,0.00}
\definecolor{orange2}{rgb}{0.93,0.60,0.00}
\definecolor{orange3}{rgb}{0.80,0.52,0.00}
\definecolor{orange4}{rgb}{0.55,0.35,0.00}
\definecolor{orangered}{rgb}{1.00,0.27,0.00}
\definecolor{orange}{rgb}{1.00,0.65,0.00}
\definecolor{orchid1}{rgb}{1.00,0.51,0.98}
\definecolor{orchid2}{rgb}{0.93,0.48,0.91}
\definecolor{orchid3}{rgb}{0.80,0.41,0.79}
\definecolor{orchid4}{rgb}{0.55,0.28,0.54}
\definecolor{orchid}{rgb}{0.85,0.44,0.84}
\definecolor{palegoldenrod}{rgb}{0.93,0.91,0.67}
\definecolor{palegreen}{rgb}{0.60,0.98,0.60}
\definecolor{paleturquoise}{rgb}{0.69,0.93,0.93}
\definecolor{paleviolet}{rgb}{0.86,0.44,0.58}
\definecolor{papayawhip}{rgb}{1.00,0.94,0.84}
\definecolor{peachpuff}{rgb}{1.00,0.85,0.73}
\definecolor{peru}{rgb}{0.80,0.52,0.25}
\definecolor{pink1}{rgb}{1.00,0.71,0.77}
\definecolor{pink2}{rgb}{0.93,0.66,0.72}
\definecolor{pink3}{rgb}{0.80,0.57,0.62}
\definecolor{pink4}{rgb}{0.55,0.39,0.42}
\definecolor{pink}{rgb}{1.00,0.75,0.80}
\definecolor{plum1}{rgb}{1.00,0.73,1.00}
\definecolor{plum2}{rgb}{0.93,0.68,0.93}
\definecolor{plum3}{rgb}{0.80,0.59,0.80}
\definecolor{plum4}{rgb}{0.55,0.40,0.55}
\definecolor{plum}{rgb}{0.87,0.63,0.87}
\definecolor{powderblue}{rgb}{0.69,0.88,0.90}
\definecolor{purple1}{rgb}{0.61,0.19,1.00}
\definecolor{purple2}{rgb}{0.57,0.17,0.93}
\definecolor{purple3}{rgb}{0.49,0.15,0.80}
\definecolor{purple4}{rgb}{0.33,0.10,0.55}
\definecolor{purple}{rgb}{0.63,0.13,0.94}
\definecolor{red1}{rgb}{1.00,0.00,0.00}
\definecolor{red2}{rgb}{0.93,0.00,0.00}
\definecolor{red3}{rgb}{0.80,0.00,0.00}
\definecolor{red4}{rgb}{0.55,0.00,0.00}
\definecolor{red}{rgb}{1.00,0.00,0.00}
\definecolor{rosybrown}{rgb}{0.74,0.56,0.56}
\definecolor{royalblue}{rgb}{0.25,0.41,0.88}
\definecolor{saddlebrown}{rgb}{0.55,0.27,0.07}
\definecolor{salmon1}{rgb}{1.00,0.55,0.41}
\definecolor{salmon2}{rgb}{0.93,0.51,0.38}
\definecolor{salmon3}{rgb}{0.80,0.44,0.33}
\definecolor{salmon4}{rgb}{0.55,0.30,0.22}
\definecolor{salmon}{rgb}{0.98,0.50,0.45}
\definecolor{sandybrown}{rgb}{0.96,0.64,0.38}
\definecolor{seagreen}{rgb}{0.18,0.55,0.34}
\definecolor{seashell1}{rgb}{1.00,0.96,0.93}
\definecolor{seashell2}{rgb}{0.93,0.90,0.87}
\definecolor{seashell3}{rgb}{0.80,0.77,0.75}
\definecolor{seashell4}{rgb}{0.55,0.53,0.51}
\definecolor{seashell}{rgb}{1.00,0.96,0.93}
\definecolor{sienna1}{rgb}{1.00,0.51,0.28}
\definecolor{sienna2}{rgb}{0.93,0.47,0.26}
\definecolor{sienna3}{rgb}{0.80,0.41,0.22}
\definecolor{sienna4}{rgb}{0.55,0.28,0.15}
\definecolor{sienna}{rgb}{0.63,0.32,0.18}
\definecolor{skyblue}{rgb}{0.53,0.81,0.92}
\definecolor{slateblue}{rgb}{0.42,0.35,0.80}
\definecolor{slategray}{rgb}{0.44,0.50,0.56}
\definecolor{slategrey}{rgb}{0.44,0.50,0.56}
\definecolor{snow1}{rgb}{1.00,0.98,0.98}
\definecolor{snow2}{rgb}{0.93,0.91,0.91}
\definecolor{snow3}{rgb}{0.80,0.79,0.79}
\definecolor{snow4}{rgb}{0.55,0.54,0.54}
\definecolor{snow}{rgb}{1.00,0.98,0.98}
\definecolor{springgreen}{rgb}{0.00,1.00,0.50}
\definecolor{steelblue}{rgb}{0.27,0.51,0.71}
\definecolor{tan1}{rgb}{1.00,0.65,0.31}
\definecolor{tan2}{rgb}{0.93,0.60,0.29}
\definecolor{tan3}{rgb}{0.80,0.52,0.25}
\definecolor{tan4}{rgb}{0.55,0.35,0.17}
\definecolor{tan}{rgb}{0.82,0.71,0.55}
\definecolor{thistle1}{rgb}{1.00,0.88,1.00}
\definecolor{thistle2}{rgb}{0.93,0.82,0.93}
\definecolor{thistle3}{rgb}{0.80,0.71,0.80}
\definecolor{thistle4}{rgb}{0.55,0.48,0.55}
\definecolor{thistle}{rgb}{0.85,0.75,0.85}
\definecolor{tomato1}{rgb}{1.00,0.39,0.28}
\definecolor{tomato2}{rgb}{0.93,0.36,0.26}
\definecolor{tomato3}{rgb}{0.80,0.31,0.22}
\definecolor{tomato4}{rgb}{0.55,0.21,0.15}
\definecolor{tomato}{rgb}{1.00,0.39,0.28}
\definecolor{turquoise1}{rgb}{0.00,0.96,1.00}
\definecolor{turquoise2}{rgb}{0.00,0.90,0.93}
\definecolor{turquoise3}{rgb}{0.00,0.77,0.80}
\definecolor{turquoise4}{rgb}{0.00,0.53,0.55}
\definecolor{turquoise}{rgb}{0.25,0.88,0.82}
\definecolor{violetred}{rgb}{0.82,0.13,0.56}
\definecolor{violet}{rgb}{0.93,0.51,0.93}
\definecolor{wheat1}{rgb}{1.00,0.91,0.73}
\definecolor{wheat2}{rgb}{0.93,0.85,0.68}
\definecolor{wheat3}{rgb}{0.80,0.73,0.59}
\definecolor{wheat4}{rgb}{0.55,0.49,0.40}
\definecolor{wheat}{rgb}{0.96,0.87,0.70}
\definecolor{whitesmoke}{rgb}{0.96,0.96,0.96}
\definecolor{white}{rgb}{1.00,1.00,1.00}
\definecolor{yellow1}{rgb}{1.00,1.00,0.00}
\definecolor{yellow2}{rgb}{0.93,0.93,0.00}
\definecolor{yellow3}{rgb}{0.80,0.80,0.00}
\definecolor{yellow4}{rgb}{0.55,0.55,0.00}
\definecolor{yellowgreen}{rgb}{0.60,0.80,0.20}
\definecolor{yellow}{rgb}{1.00,1.00,0.00}

\definecolor{white1}{rgb}{1.00,1.00,1.00}

\definecolor{frontal1}{rgb}{0.93,0.27,0.21}
\definecolor{lateral1}{rgb}{0.68,0.66,0.80}
\definecolor{horizontal1}{rgb}{0.94,0.63,0.37}

\definecolor{tnode1}{rgb}{0.93,0.27,0.21}
\definecolor{mnode1}{rgb}{0.68,0.66,0.80}

\definecolor{unsure1}{rgb}{1.00,0.00,0.46}

\graphicspath{{01-figures/}{02-tikz/}}

\title{HOTTBOX: Higher Order Tensors ToolBOX}
\author[1]{Ilya Kisil}
\author[1]{Giuseppe G. Calvi}
\author[1]{Bruno S. Dees}
\author[1]{Danilo P. Mandic}
\affil[1]{
    \small
    Department of Electrical and Electronic Engineering, Imperial College London, SW7 2AZ, UK\\
    E-mails: \{ik1614, ggc115, bs1912, d.mandic\}@imperial.ac.uk
}

\newcommand{\citep}[1]{{\cite{#1}}}
\usepackage{amssymb}
\usepackage{amsmath}
\usepackage{stmaryrd}       
\usepackage{mathcomp}
\usepackage[capitalise]{cleveref}
\usepackage{amsthm}
\newtheorem*{remark}{Remark}

\begin{document}

\title{HOTTBOX: Higher Order Tensor ToolBOX for the Analysis of Multi-way Data}

\maketitle

\begin{abstract}%
    HOTTBOX is a Python library for exploratory analysis and visualisation of
    multi-dimensional arrays of data, also known as tensors. The library includes methods ranging from standard
    multi-way operations and data manipulation through to multi-linear algebra based
    tensor decompositions. HOTTBOX also comprises
    sophisticated algorithms for generalised multi-linear classification and data fusion,
    such as Support Tensor Machine (STM) and Tensor Ensemble Learning (TEL).
    For user convenience, HOTTBOX offers a unifying API which establishes a
    self-sufficient ecosystem for various forms of efficient representation of multi-way
    data and the corresponding decomposition and association algorithms. Particular emphasis
    is placed on scalability and interactive visualisation, to support multidisciplinary data analysis
    communities working on big data and tensors. HOTTBOX also provides means for
    integration with other  popular data science
    libraries for visualisation and data manipulation. The source code, examples and
    documentation ca be found at \texttt{\url{https://github.com/hottbox/hottbox}}.
\end{abstract}


\section{Introduction}
Tensors are higher order generalisations of matrices and vectors, which represent
data in the form of a multi-way array. This is achieved through indexing by an arbitrary number of
indices (tensor order), whereby different physically meaningful properties and characteristics are
ideally
attributed to different dimensions (modes) of an N-dimensional array.  Such data representation
 offers a whole host of opportunities for discovering
 latent dependencies and intrinsic structures in data.
Owing to the inherent flexibility and horizontal scalability of multi-way analysis,
tensors have found application in a diverse range of disciplines, from very theoretical
ones, such as physics and numerical analysis
\citep{de2006link, cichocki2017tensor},
\red{
    to the more applied areas, such as face
    recognition \citep{vasilescu2002multilinear}, separation of unknown sources of brain
    activity \citep{morup2011applications}, cross-language information retrieval
    \citep{chew2007cross} and computational neuroscience \citep{sen2017extraction,
    wang2018extracting}, to mention but a few.
}
However, both the analysis and wider adoption
of tensors are still being hampered by a lack of software tools which offer interactive
scientific visualisation throughout the analysis.

Over the past decade, extensive literature has been published on multi-way analysis,
owing to the ability of tensor decompositions to alleviate the ``Curse of
Dimensionality'', such techniques range from
fundamental principles behind tensor decompositions \citep{kolda2009tensor, cichocki2015tensor} to quite
advanced and sophisticated techniques such as tensor networks
\citep{cichocki2016tensor}.
In addition, we have witnessed a rapid growth of the amount of
data being generated and  collected, together with the corresponding storage capabilities and processing power,
these are readily available through either local or cloud computing to users even outside
of academia and industry.  The multi-way analysis community has responded with the
development of open source packages \citep{software-tensor-toolbox, software-tensorlab,
software-tensorly, software-tensord} in order to mitigate, for a generally
knowledgeable user, most of the technicalities behind implementation of tensor
decompositions, alongside the corresponding statistical methods and data analysis
techniques. Existing publicly available libraries are mainly designed as
general purpose toolboxes comprising fundamental tensor decompositions, e.g.
Canonical Polyadic Decomposition (CPD)
, Tucker Decomposition (TKD) 
and Tensor Train Decomposition (TT)
, with a focus on efficient performance of such methods.


The Higher Order Tensor ToolBOX (HOTTBOX), presented here, is an open source
toolbox for tensor decompositions, statistical analysis, visualisation, feature
extraction, regression and non-linear classification of multi-dimensional data.  The
package is written in Python and was conceived with the aim to serve as a self-contained
library which allows for seamless integration with other popular data science packages
for data wrangling, together with offering unique visualisation capability suitable for
non-experts and multidisciplinary research communities.
The HOTTBOX has already been rolled out as a research tool and for
educational purposes \citep{moniri2018refreshing} and it is our hope that it will both help
demystify tensor decompositions for various communities active in the area, and also
attract the curious reader to gain experience with multi-way analysis.

\section{Background: Available software for multi-linear operations}
Over the past years, several libraries for dealing with multi-dimensional data have
emerged, including:

\begin{itemize}
    \item \textit{Tensor Toolbox} \citep{software-tensor-toolbox}, one of the first
        attempts to provide users with classes and functions for manipulating dense,
        sparse and structured tensors as well as fundamental tensor decompositions using
        MATLAB's object-oriented features;

    \item \textit{Tensorlab} \citep{software-tensorlab}, which
        offers various tools for tensor computations and complex optimisations.  In
        addition, it also implements a number of algorithms dealing with large-scale
        datasets and more advanced factorisations, such as the LL1 and Block Term
        Decomposition (BTD) \citep{sorber2013optimization, de2008decompositions}, and
        supports quite intuitive tools for basic visualisations of N-dimensional data.
\end{itemize}

Both these toolboxes are well structured and provide extensive documentation supported
with various examples; they also require a proprietary MATLAB platform. Existing packages
written in Python include:

\begin{itemize}
    \item \textit{TensorLy} \citep{software-tensorly},  which was developed with
        simplicity in mind and has a
        flexible backend system which allows to perform tensor decomposition algorithms
        at scale and over a range of hardware setups.  TensorLy also comprises implementations
        of some machine learning approaches, but its main focus remains on optimised and
        efficient performance;

    \item The \textit{TensorD} toolbox \citep{software-tensord} was designed as a highly
        modular piece of software, specifically tailored to the most prominent machine
        learning framework, the TensorFlow.  Although it allows us to transform an idea
        into a result very quickly, it supports implementations of only CPD and TKD.
\end{itemize}

Several other libraries exist which primary focus on the Tensor Train Decomposition
(TTD), a tensor factorisation originally introduced by a research group from the Skolkovo
Institute of Science and Technology in 2011.  The same group also released a Python
implementation of their algorithms within a dedicated toolbox
whose functionalities were then extended to support utilisation of hardware acceleration
for efficient computations, batch processing and automatic differentiation
\citep{software-t3f}.
Another package for tensor network learning with PyTorch
takes a
somewhat orthogonal approach to establishing a common interface,  where the existing
multi-linear models, i.e. CPD, TKD, TT, are represented as a particular case of tensor
networks \citep{cichocki2016tensor}.  Thus, decomposing, manipulating, and reconstructing
tensors can be (to some extent) abstracted away from the particular decomposition format.


To summarise, while all of the aforementioned libraries and projects have found their place
within multi-way data analysis, the HOTTBOX takes a step further by:
\begin{itemize}
    \item Utilising tensors not only as N-dimensional arrays, but also exploiting
        complementary meta-information about the data a tensor represents;

    \item Offering scalable visualisation tools, at every stage of the analysis, to
        facilitate ease of analysis, multi-disciplinary efforts or educational purposes;

    \item Being focused on unifying the machine learning approaches which
        are specific to inherently multi-dimensional data.

\end{itemize}

\begin{figure}[!tb]
    \includegraphics[width=\linewidth]{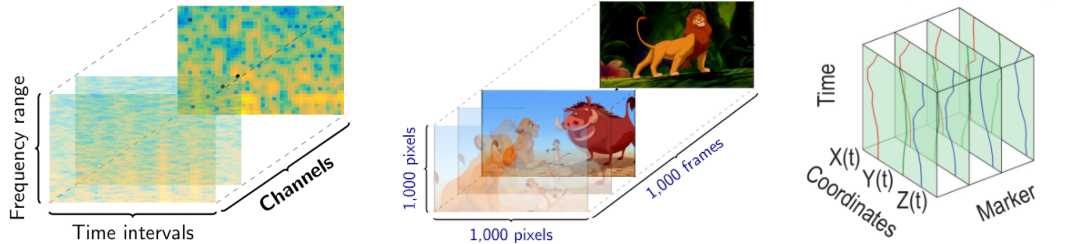}
    \centering
    \caption{
        Biomedical data (left), a video sequence (middle), and 3-D motion capturing
        system (right) organised as ``raw'' multi-dimensional data.
    }
    \label{fig:tensors}
\end{figure}
\section{Ecosystem of HOTTBOX}

Various types of data can be represented in the form of an N-dimensional array.  Most
often, this is achieved through a combination of mathematical construction and
experimental design.  For example, tensors can be constructed from:
\begin{itemize}
    \item Time-frequency representation of each channel that records brain activity
        during a particular task or stimulus, that is, each slice on the left-most panel in
        \cref{fig:tensors} \red{\citep{cong2015tensor,zhou2016linked}};

    \item A video clip can be broken down into a set of consecutive frames which are
        stacked into a tensor, as in the middle panel in \cref{fig:tensors}
        \red{\citep{sobral2015online, jiang2018no}};

    \item Optical motion capture system that contains several markers, each of which can
        record information about position change in a 3-D space over a period of time, as
        illustrated in the right most-panel in \cref{fig:tensors}
        \red{\citep{kruger2008multi,hou2014scalable}}.
\end{itemize}

\subsection{Core structural components of HOTTBOX}\label{sec:core-components}

\begin{figure}[!tb]
    \centering
    \begin{minipage}{0.45\textwidth}
        \centering
        \includegraphics[width=0.9\linewidth, height=0.25\textheight]{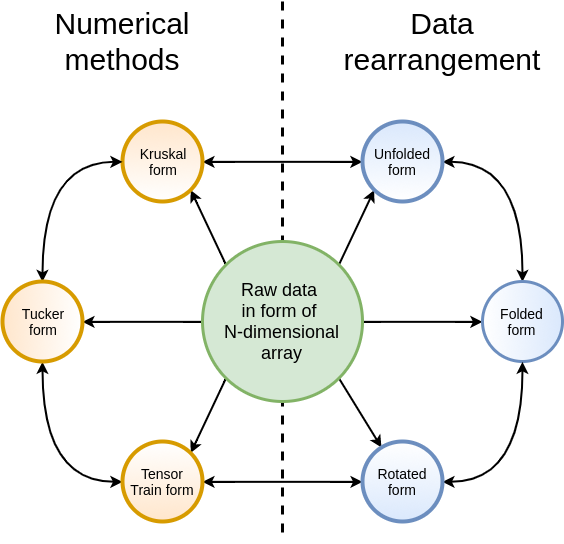}
        \label{fig:different-forms-of-data}
    \end{minipage}%
    \begin{minipage}{.55\textwidth}
        \centering
        \includegraphics[width=0.9\linewidth, height=0.25\textheight]{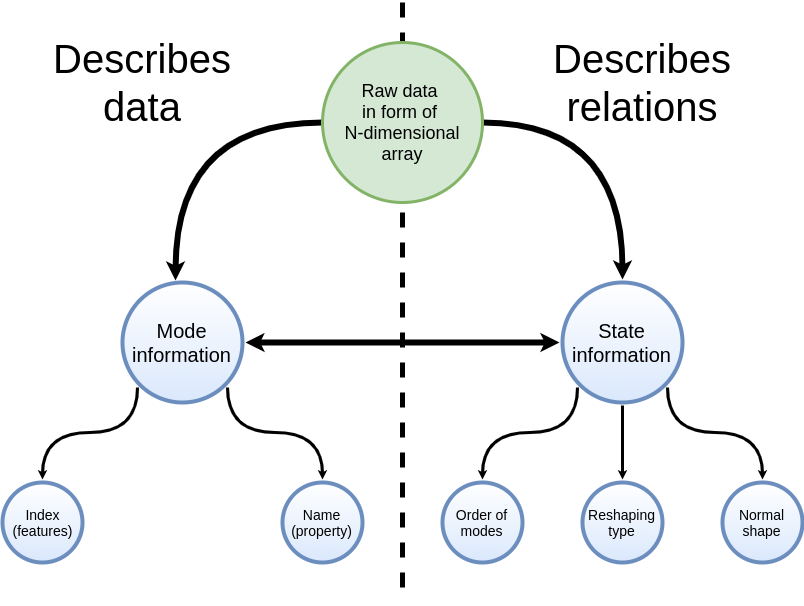}
        \label{fig:data-modes-state}
    \end{minipage}
    \caption{
        Data and metadata formals within the HOTTBOX.
        Left: The ecosystem of representations of an N-dimensional array.
        Right: The synergy of meta-information associated with an N-dimensional array
        descriptions.
    }
\end{figure}

Each dimension within an N-dimensional array can be associated with a certain property,
or mode of the raw data.  At the same time, this characteristic is described by a set of
distinct features.  To show this, assume that a simple data rearrangement procedure (e.g.
folding, unfolding) of the raw data was performed which yielded a different view of data
at hand.  However, such manipulation does not change the original properties of the
underlying data array, but instead it changes the relationships between models. We shall
refer to such transformation of a ``raw'' N-dimensional data as a \textit{change of
state}. The HOTTBOX ecosystem therefore establishes the following principles and objects:

\begin{itemize}
    \item \textit{Mode} of the tensor is defined by the name of the property it
        represents and names of the features that describe this property;

    \item \textit{State} of the tensor is defined by transformations applied to the data
        array;

    \item \textit{Meta-data} about a tensor is a record of information about the modes
        and the state it is currently in.

\end{itemize}

Oftentimes, N-dimensional data contains a considerable amount of repetitive and redundant
information.  By applying different numerical methods, i.e. tensor decompositions, this
information can be represented in a more compact and efficient way.  There are three
major conceptually different factorisation types and computational algorithms associated
with tensors that are part of HOTTBOX:
\begin{itemize}
    \item \textit{Kruskal form} represents the original data as a list of matrices, each
        of which corresponds to a particular dimension of an array.  It also imposes a
        one-to-one relation between column vectors of these factor matrices;

    \item \textit{Tucker form} is similar to Kruskal form, but every vector within  one factor
        matrix is related to all vectors of the other factor matrices through a core
        tensor;

    \item \textit{Tensor Train form} represents a tensor as a set of sparsely
        interconnected matrices and core tensors of low order, and therefore only
        adjacent components have explicit relation.

\end{itemize}

It is important to understand that the characteristics of modes of N-dimensional data
remain the same regardless of the form used for their representation. Therefore, to suit the
inherent characteristics of tensors, HOTTBOX provides a unified API which is very
convenient as it:
\begin{itemize}
    \item Allows for switching between different tensor forms seamlessly;

    \item Preserves information about the underlying tensor characteristics;

    \item Gives an opportunity to utilise meta-information with ease.
\end{itemize}

The core structural components of HOTTBOX are therefore the \texttt{Tensor},
\texttt{TensorCPD}, \texttt{TensorTKD} and \texttt{TensorTT} classes which bring together
the raw, Kruskal, Tucker, and Tensor Train formats, respectively.

\subsection{Supported algorithms}
\begin{figure}[tb]
    \includegraphics[width=.85\linewidth]{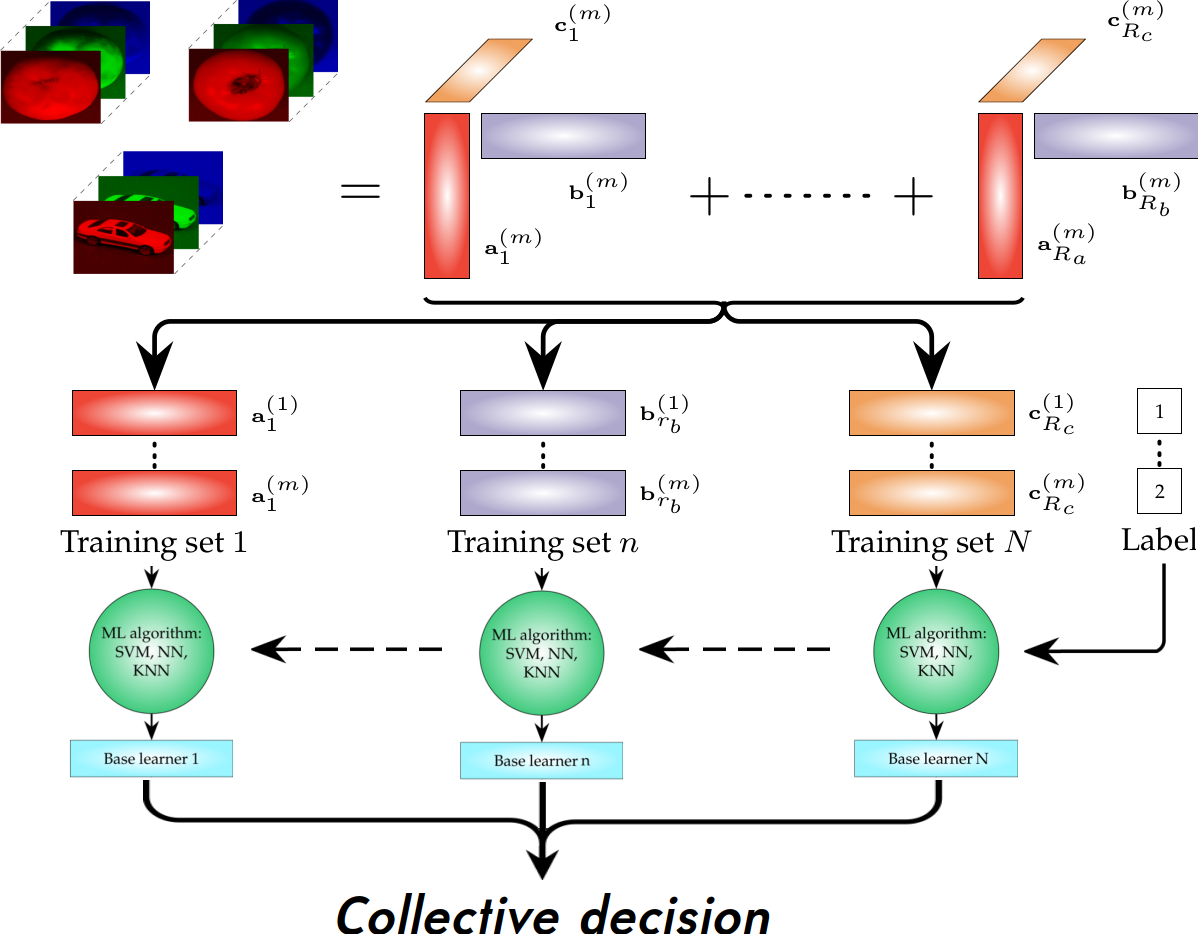}
    \centering
    \caption{
        A conceptual diagram of the \textit{Tensor Ensemble Learning} (TEL) framework
        within HOTTBOX.
        Each N-dimensional data sample (top left) from the original dataset is
        represented through either \texttt{TensorCPD} or \texttt{TensorTKD} (top right).
        The so-extracted components are reorganised and form a series of new datasets,
        each containing an incomplete information about the original sample (middle). An
        ensemble of machine learning algorithms is then employed, whereby each base
        learner generates an independent hypothesis which then used for ``voting'' in
        order to form a collective decision (bottom).
    }
    \label{fig:tel-concept}
\end{figure}

One aspect of the HOTTBOX library is to facilitate the usage of a wide range of existing
numerical methods  which are available for analysis of inherently multi-dimensional data.
The HOTTBOX splits implementation of such algorithms into several modules dedicated to
different aspects of multi-way analysis, as follows:

\begin{itemize}
    \item The \textit{tensor factorisation module} comprises of implementations of
        fundamental tensor decompositions which include the Canonical Polyadic, Higher
        Order Singular Value and Tensor Train  decompositions
        \citep{kolda2009tensor,de2000multilinear,oseledets2011tensor} that can be used to
        represent a ``raw'' multi-modal data as one of the structures covered in
        \cref{sec:core-components}.  A set of these algorithms is further extended with
        different variations which have been proposed in the literature on optimised
        processing of multi-dimensional arrays.  For example, the Higher Order Orthogonal
        Iteration (HOOI) Decomposition \citep{de2000best} aids the actual computation of
        the best multi-linear approximation, while the Randomised Canonical Polyadic
        Decomposition \citep{battaglino2018practical} significantly reduces both memory
        requirements and computational speed, without sacrificing precision of the CP
        factorisation;

    \item Inside the \textit{data fusion and feature extraction} module, the user can
        find the Coupled Matrix and Tensor factorisation \citep{jeon2016scout} and
        PARAFAC2 \citep{kiers1999parafac2} routines. Both can be employed to jointly analyse a
        multi-way array with a matrix or a collection of matrices that share a common
        mode (which describes similar data properties) in order to extract a latent
        structure that governs the underlying processes. Despite the latter model
        being able to simultaneously factorise a collection of only 2-dimensional
        structures, PARAFAC2 is still attributed to multi-way analysis as a variant of
        the CP representation with relaxed constrains;

    \item The \textit{classification and regression} module includes implementations of
        the Least Squares Support Tensor Machine (LSSTM) and Tensor Ensemble Learning
        (TEL) algorithms (illustrated in \cref{fig:tel-concept}) which respectively
        generalise concepts and ideas of the existing popular ``flat-view'' matrix
        frameworks, i.e. Support Vector Machine and Ensemble Learning frameworks
        \citep{cichocki2016tensor,kisil2018tensor} to their tensor counterparts, such as
        Support Tensor Machine (STM).  Thus, the LSSTM and TEL can be seen as a natural
        extension to the case of tensor-valued data samples.

\end{itemize}

The HOTTBOX is designed with the aim to offer easy integration of other emerging ML
routines for Big Data.

\subsection{Interactive visualisations}
\begin{figure}[!tb]
    \centering
    \includegraphics[width=\linewidth]{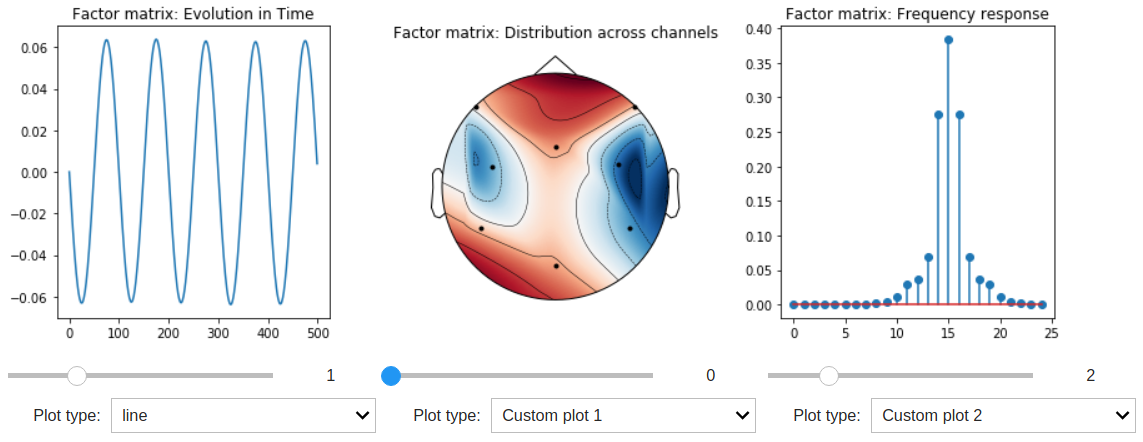}
    \caption{
        An example of a dashboard for visualising the \texttt{TensorTKD}. It fully
        exploits the relationships between components and utilises the associated
        meta-information.
    }
    \label{fig:visualisation-1}
\end{figure}
A key benefit of having a common API in our HOTTBOX library is that it provides us with
the means to integrate the multi-linear structures with external popular libraries that
serve completely orthogonal purposes, e.g. data wrangling and visualisation. The latter
is particularly important for exploratory data analysis, and when displaying various
forms of tensor representations. This is completely different from commonly used
toolboxes, as current approaches require a considerable amount of repetitive work to
achieve an acceptable and effective visualisation result.  To this end, in order to
accommodate many versatile forms of efficient representations, the HOTTBOX provides users
with an interactive dashboard. \cref{fig:visualisation-1} illustrates such a
visualisation with the
user interface for a ``raw'' N-dimensional array factorised into the
Tucker form. Due to the nature of this representation, different modes and components
correspond to different properties that can be plotted using various types of graphs
available in the dropdown menu. Although a set of options is predefined, it can be easily
extended to suit a particular need of a user. In this way, a typically very large number of
cross-modal relations intrinsic to the components of Tucker form (or other forms of
efficient representations) can be explored with the use of sliders. Finally,
meta-information is extracted and embedded as figure titles, to keep track of the design
sequence.

\red{
    \begin{remark}
        \cref{fig:visualisation-1} merely displays synthetic data for illustrational
        purposes, where each panel visualises a single column vector from three different
        factor matrices that would have been obtained through either CPD or TKD of
        brain activity represented in the form of a third order tensor.
    \end{remark}
}

\subsection{Statistical toolbox}
\begin{figure}[tb]
    \centering
    \includegraphics[width=0.65\textwidth, trim={0.2cm 0cm 0 0cm},clip]{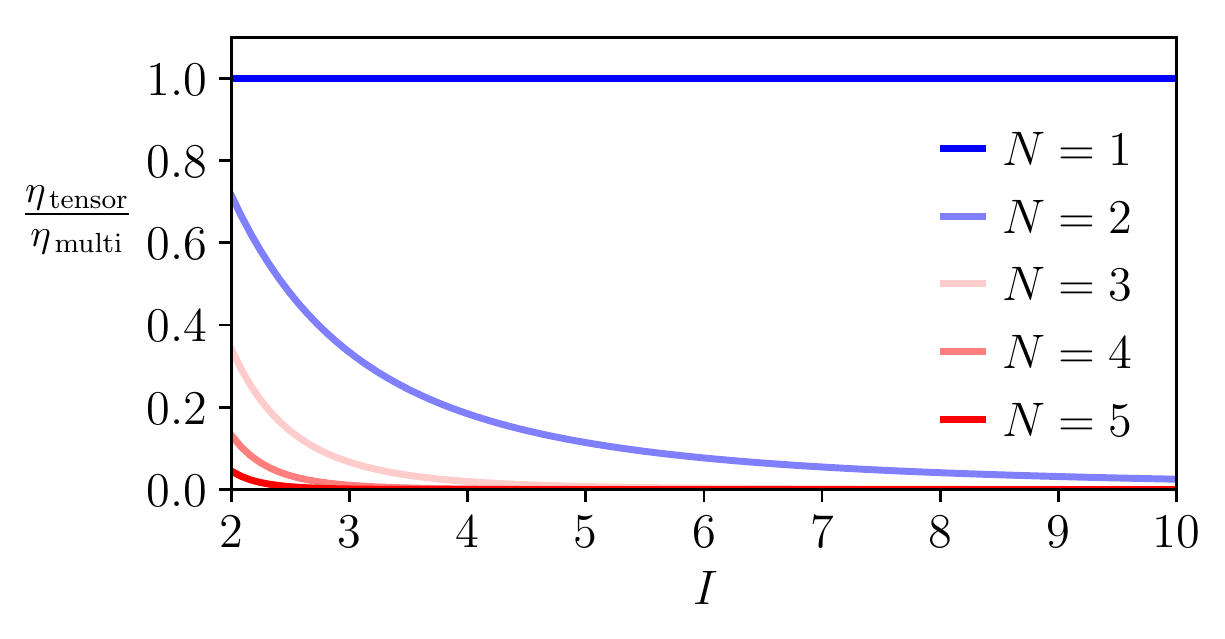}
    \vspace{-0.4cm}
    \caption{
        The ratio of distinct parameters between the structured tensor model,
        $\eta_{\,\text{tensor}}$, and the classical unstructured model,
        $\eta_{\,\text{multi}}$, given by
        $\frac{\eta_{\,\text{tensor}}}{\eta_{\,\text{multi}}}$, for a varying mode
        dimensionality, $I$, and tensor order, $N$.
    }
    \label{fig:dof}
\end{figure}
The HOTTBOX also provides classes and functions
for the estimation of \textit{Tensor-Valued Gaussian} statistical models
.
The inherent notion of tensor-valued probability density function
makes this module  suitable for a large variety of use-cases that aim to model
interactions between real-world tensor data samples that are typically causal and
probabilistic. Our implementation is compact, stemming from the Kronecker separability of the
first- and second-order moments of the tensor-valued random variable, as elaborated in
\citep{Scalzo2019_1}. This property of tensor-valued statistical models provides
an immediate reduction in the number of parameters required to represent a random process compared to
the standard multivariate Gaussian model, as can be observed in \cref{fig:dof}.

The so-enabled exploitation of tensor-valued models is essential for alleviating the
\textit{Curse of Dimensionality}, that is, an exponential growth in data volume with an
increase in the order of the tensor. At the same time, implementation of tensor-valued
Gaussian statistical models within HOTTBOX equips researches and data analysts with the
essential tool for:
\begin{itemize}
    \item Utilising Bayesian tensor inference methods;

    \item Straightforward consideration of a mixture of probabilistic models;

    \item Statistical and hypothesis testing through the tensor likelihood function;

    \item Employing class-conditional densities for classification tasks.
\end{itemize}

\section{Conclusion and future work}
The Higher Order Tensor ToolBOX (HOTTBOX) is a Python library dedicated to handling
multi-dimensional data, which at its core employs a user-friendly approach to
statistical analysis, visualisation, feature extraction, regression and non-linear
classification paradigms.  In this way, HOTTBOX can be used both by the data
analytics communities and the multi-disciplinary communities which operate with
inherently N-dimensional data arrays.  The toolbox is equipped with a unit tests suite in
order to meet the requirements on the integrity of the code base of the modern
software.  A particular emphasis has been placed on the importance of
documentation\footnote{\url{https://hottbox.github.io/develop}} which is automatically
updated to reflect the most recent development changes and to include the latest
features.  An interested reader can find more details in our publicly available repositories%
\footnote{\url{https://github.com/hottbox/hottbox-tutorials}} which contain a series of
illustrative examples that can be tried out live in a cloud, for free and without any
installation burden.

The ease of use makes the HOTTBOX also suitable for educational purposes, as a means to
shed the light on the field of multi-linear analysis. For example, it has been used to
facilitate the final year undergraduate and post graduate curriculum%
\footnote{\url{https://github.com/IlyaKisil/dpm-coursework}}
within the courses ``Adaptive Signal Processing and Machine Intelligence'' and ``Machine
Learning for Finance'', at the authors' home institution, enabling a seamless transition
from the flat-view linear algebra to multi-linear analysis of tensors.




\newpage
\bibliographystyle{IEEE}
\bibliography{%
    04-bibliography/references,%
    04-bibliography/references-applications,%
    04-bibliography/references-software,%
    04-bibliography/references-BSD%
}

\begin{thebibliography}{10}

\bibitem{de2006link}
L.~De~Lathauwer,
\newblock ``A link between the canonical decomposition in multilinear algebra
  and simultaneous matrix diagonalization,''
\newblock {\em SIAM journal on Matrix Analysis and Applications}, vol. 28, no.
  3, pp. 642--666, 2006.

\bibitem{cichocki2017tensor}
A.~Cichocki, A.~H. Phan, Q.~Zhao, N.~Lee, I.~Oseledets, M.~Sugiyama, and D.~P.
  Mandic,
\newblock ``Tensor networks for dimensionality reduction and large-scale
  optimization. {{Part}} 2: {{Applications}} and future perspectives,''
\newblock {\em Foundations and Trends in Machine Learning}, vol. 9, no. 6, pp.
  431--673, 2017.

\bibitem{vasilescu2002multilinear}
M.~A.~O. Vasilescu and D.~Terzopoulos,
\newblock ``Multilinear analysis of image ensembles: Tensorfaces,''
\newblock {\em In Proceedings of the European Conference on Computer Vision},
  pp. 447--460, 2002.

\bibitem{morup2011applications}
M.~M{\o}rup,
\newblock ``Applications of tensor (multiway array) factorizations and
  decompositions in data mining,''
\newblock {\em Wiley Interdisciplinary Reviews: Data Mining and Knowledge
  Discovery}, vol. 1, no. 1, pp. 24--40, 2011.

\bibitem{chew2007cross}
P.~A. Chew, B.~W. Bader, T.~G. Kolda, and A.~Abdelali,
\newblock ``{Cross-language information retrieval using PARAFAC2},''
\newblock {\em In Proceedings of the 13th International Conference on Knowledge
  Discovery and Data Mining}, pp. 143--152, 2007.

\bibitem{sen2017extraction}
B.~Sen and K.~K. Parhi,
\newblock ``{Extraction of common task signals and spatial maps from group fMRI
  using a PARAFAC-based tensor decomposition technique},''
\newblock {\em {In Proceedings of the IEEE International Conference on
  Acoustics, Speech and Signal Processing (ICASSP)}}, pp. 1113--1117, 2017.

\bibitem{wang2018extracting}
Deqing Wang, Yongjie Zhu, Tapani Ristaniemi, and Fengyu Cong,
\newblock ``{Extracting multi-mode ERP features using fifth-order nonnegative
  tensor decomposition},''
\newblock {\em Journal of neuroscience methods}, vol. 308, pp. 240--247, 2018.

\bibitem{kolda2009tensor}
T.~G. Kolda and B.~W. Bader,
\newblock ``Tensor decompositions and applications,''
\newblock {\em SIAM Review}, vol. 51, no. 3, pp. 455--500, 2009.

\bibitem{cichocki2015tensor}
A.~Cichocki, D.~P. Mandic, L.~De~Lathauwer, G.~Zhou, Q.~Zhao, C.~Caiafa, and
  H.~A. Phan,
\newblock ``Tensor decompositions for signal processing applications: From
  two-way to multiway component analysis,''
\newblock {\em IEEE Signal Processing Magazine}, vol. 32, no. 2, pp. 145--163,
  2015.

\bibitem{cichocki2016tensor}
A.~Cichocki, N.~Lee, I.~Oseledets, A.~H. Phan, Q.~Zhao, and D.~P. Mandic,
\newblock ``Tensor networks for dimensionality reduction and large-scale
  optimization. {{Part}} 1: {{Low-rank}} tensor decompositions,''
\newblock {\em Foundations and Trends in Machine Learning}, vol. 9, no. 4-5,
  pp. 249--429, 2016.

\bibitem{software-tensor-toolbox}
B.~W. Bader, T.~G. Kolda, and \textit{et al},
\newblock ``{MATLAB Tensor Toolbox Version 3.1},'' .

\bibitem{software-tensorlab}
N.~Vervliet, O.~Debals, and L.~De~Lathauwer,
\newblock ``{Tensorlab 3.0 -- Numerical optimization strategies for large-scale
  constrained and coupled matrix/tensor factorization},''
\newblock {\em In Proceedings of the 50th Asilomar Conference on Signals,
  Systems and Computers}, pp. 1733--1738, 2016.

\bibitem{software-tensorly}
J.~Kossaifi, Y.~Panagakis, A.~Anandkumar, and M.~Pantic,
\newblock ``{Tensorly: Tensor learning in Python},''
\newblock {\em The Journal of Machine Learning Research}, vol. 20, no. 1, pp.
  925--930, 2019.

\bibitem{software-tensord}
L.~Hao, S.~Liang, J.~Ye, and Z.~Xu,
\newblock ``{TensorD: A tensor decomposition library in TensorFlow},''
\newblock {\em Neurocomputing}, vol. 318, pp. 196--200, 2018.

\bibitem{moniri2018refreshing}
A.~Moniri, I.~Kisil, A.~G. Constantinides, and D.~P. Mandic,
\newblock ``{Refreshing DSP courses through biopresence in the curriculum: A
  successful paradigm},''
\newblock {\em In Proceedings of the 23rd IEEE International Conference on
  Digital Signal Processing (DSP)}, pp. 1--4, 2018.

\bibitem{sorber2013optimization}
L.~Sorber, M.~Van~Barel, and L.~De~Lathauwer,
\newblock ``{Optimization-based algorithms for tensor decompositions: Canonical
  polyadic decomposition, decomposition in rank-$(L_r,L_r,1)$ terms, and a new
  generalization},''
\newblock {\em SIAM Journal on Optimization}, vol. 23, no. 2, pp. 695--720,
  2013.

\bibitem{de2008decompositions}
L.~De~Lathauwer,
\newblock ``Decompositions of a higher-order tensor in block terms. {{Part}}
  {{II}}: Definitions and uniqueness,''
\newblock {\em SIAM Journal on Matrix Analysis and Applications}, vol. 30, no.
  3, pp. 1033--1066, 2008.

\bibitem{software-t3f}
A.~Novikov, P.~Izmailov, V.~Khrulkov, M.~Figurnov, and I.~Oseledets,
\newblock ``{Tensor Train decomposition on TensorFlow (T3F)},''
\newblock {\em arXiv preprint arXiv:1801.01928}, 2018.

\bibitem{cong2015tensor}
F.~Cong, Q.~Lin, L.~Kuang, X.~Gong, P.~Astikainen, and T.~Ristaniemi,
\newblock ``{Tensor decomposition of EEG signals: a brief review},''
\newblock {\em Journal of Neuroscience Methods}, vol. 248, pp. 59--69, 2015.

\bibitem{zhou2016linked}
G.~Zhou, Q.~Zhao, Y.~Zhang, T.~Adal{\i}, S.~Xie, and A.~Cichocki,
\newblock ``{Linked component analysis from matrices to high-order tensors:
  Applications to biomedical data},''
\newblock {\em Proceedings of the IEEE}, vol. 104, no. 2, pp. 310--331, 2016.

\bibitem{sobral2015online}
S.~Andrews, J.~Sajid, K.~J. Soon, B.~Thierry, and Z.~El-hadi,
\newblock ``Online stochastic tensor decomposition for background subtraction
  in multispectral video sequences,''
\newblock {\em In Proceedings of the IEEE International Conference on Computer
  Vision Workshops}, pp. 106--113, 2015.

\bibitem{jiang2018no}
G.~Jiang, S.~Liu, M.~Yu, F.~Shao, Z.~Peng, and F.~Chen,
\newblock ``No reference stereo video quality assessment based on motion
  feature in tensor decomposition domain,''
\newblock {\em Journal of Visual Communication and Image Representation}, vol.
  50, pp. 247--262, 2018.

\bibitem{kruger2008multi}
B.~Kr{\"u}ger, J.~Tautges, M.~M{\"u}ller, and A.~Weber,
\newblock ``Multi-mode tensor representation of motion data,''
\newblock {\em Journal of Virtual Reality and Broadcasting}, vol. 5, no. 5,
  2008.

\bibitem{hou2014scalable}
J.~Hou, L.~Chau, N.~Magnenat-Thalmann, and Y.~He,
\newblock ``Scalable and compact representation for motion capture data using
  tensor decomposition,''
\newblock {\em IEEE Signal Processing Letters}, vol. 21, no. 3, pp. 255--259,
  2014.

\bibitem{de2000multilinear}
L.~De~Lathauwer, B.~De~Moor, and J.~Vandewalle,
\newblock ``A multilinear singular value decomposition,''
\newblock {\em SIAM Journal on Matrix Analysis and Applications}, vol. 21, no.
  4, pp. 1253--1278, 2000.

\bibitem{oseledets2011tensor}
I.~Oseledets,
\newblock ``Tensor-train decomposition,''
\newblock {\em SIAM Journal on Scientific Computing}, vol. 33, no. 5, pp.
  2295--2317, 2011.

\bibitem{de2000best}
L.~De~Lathauwer, B.~De~Moor, and J.~Vandewalle,
\newblock ``On the best rank-1 and rank-$(r_1, r_2,..., r_n)$ approximation of
  higher-order tensors,''
\newblock {\em SIAM journal on Matrix Analysis and Applications}, vol. 21, no.
  4, pp. 1324--1342, 2000.

\bibitem{battaglino2018practical}
C.~Battaglino, G.~Ballard, and T.~G. Kolda,
\newblock ``{A practical randomized CP tensor decomposition},''
\newblock {\em SIAM Journal on Matrix Analysis and Applications}, vol. 39, no.
  2, pp. 876--901, 2018.

\bibitem{jeon2016scout}
B.~Jeon, I.~Jeon, L.~Sael, and U.~Kang,
\newblock ``{Scout: Scalable coupled matrix-tensor factorization-algorithm and
  discoveries},''
\newblock {\em In Proceedings of the 32nd IEEE International Conference on Data
  Engineering (ICDE)}, pp. 811--822, 2016.

\bibitem{kiers1999parafac2}
H.~A.~L. Kiers, J.~M.~F. Ten~Berge, and R.~Bro,
\newblock ``{PARAFAC2 Part I: A direct fitting algorithm for the PARAFAC2
  model},''
\newblock {\em Journal of Chemometrics}, vol. 13, no. 3-4, pp. 275--294, 1999.

\bibitem{kisil2018tensor}
I.~Kisil, A.~Moniri, and D.~P. Mandic,
\newblock ``Tensor ensemble learning for multidimensional data,''
\newblock {\em In Proceedings for the IEEE Global Conference on Signal and
  Information Processing (GlobalSIP)}, pp. 1358--1362, 2018.

\bibitem{Scalzo2019_1}
B.~Scalzo~Dees and D.~P. Mandic,
\newblock ``{A statistically identifiable model for tensor-valued Gaussian
  random variables},''
\newblock {\em {arXiv:1911.02915}}, 2019.

\end{thebibliography}

\end{document}